\documentclass[twocolumn,eqsecnum,aps,showpacs]{revtex4}

\begin{document}   
\title{Effect of quantum fluctuations on topological excitations 

and central charge in supersymmetric theories}
\author{K. Shizuya}
\affiliation{Yukawa Institute for Theoretical Physics\\
Kyoto University,~Kyoto 606-8502,~Japan }

\begin{abstract} 
The effect of quantum fluctuations on Bogomol'nyi-Prasad-Sommerfield
(BPS)-saturated topological excitations  in supersymmetric theories is studied.
Focus is placed on a sequence of topological excitations that derive 
from the same classical soliton or vortex in lower dimensions and 
it is shown that their quantum characteristics, such as the spectrum 
and profile, differ critically with the dimension of spacetime. 
In all the examples examined the supercharge algebra retains its
classical form although short-wavelength fluctuations may modify
the operator structure of the central charge, yielding an anomaly.
The central charge, on taking the expectation value, is further
affected by long-wavelength fluctuations, and this makes the
BPS-excitation spectra only approximately calculable in some
low-dimensional theories.  In four dimensions, in contrast, holomorphy
plays a special role  in stabilizing the BPS-excitation spectra against
quantum corrections.    The basic tool in our study is the superfield
supercurrent,  from which the supercharge algebra with a central extension
is  extracted in a supersymmetric setting.
A general method is developed to determine the associated 
superconformal anomaly by considering dilatation directly in superspace.

\end{abstract}

\pacs{11.10.Kk, 11.30.Pb}

\maketitle

\section{Introduction}

There has recently been renewed interest in topological excitations
in supersymmetric theories, 
in connection with the brane-world scenarios~\cite{DS} and
nonperturbative  analysis of supersymmetric theories~\cite{SW}. In
particular, topological excitations that saturate the
Bogomol'nyi-Prasad-Sommerfield (BPS) bound~\cite{B} provide a mechanism
for spontaneous supersymmetry breaking,  with the world localized on a
topological defect preserving only part of the original supersymmetry.

In the context of supersymmetry BPS saturation has a special
meaning~\cite{WO,W}:
Normally BPS-saturated excitations belong to a shorter supermultiplet 
and remain saturated once they are so classically; 
their spectrum thereby is related to the central charge in the supercharge algebra.  
This, of course, does not mean that the spectrum stays classical.
Indeed it was clarified after years of
investigation~\cite{Schon,RN,NSNR,GJ,SVV,LSV,RNW,FN,GRNW} 
that solitons in a two-dimensional (2d) theory with N=1 supersymmetry
remain BPS saturated while the associated central charge acquiring 
an anomaly at the quantum level~\cite{SVV}.

Topological excitations, such as solitons, domain walls, junctions and
vortices, have long been studied in a variety of contexts and
in recent years some interesting classical multi-body configurations 
of topological excitations~\cite{TIT}
have also been explored in the context of supersymmetry. 
It would be of importance to
consider how such classical excitations of interest behave
against quantum fluctuations.

The purpose of this paper is to study the effect of quantum
fluctuations on BPS-saturated topological excitations 
in supersymmetric theories.  
Focus is placed on a sequence of topological excitations that derive from
the same classical soliton or vortex in lower dimensions and it is shown 
that their quantum characteristics, such as the spectrum and profile, 
differ critically with the dimension of spacetime. 
Also pointed out is a special role played by holomorphy inherent for 4d
supersymmetry in stabilizing the BPS-excitation spectra against quantum
corrections. 

The basic tool in our study is the superfield supercurrent~\cite{FZ}, 
from which the supercharge algebra with a central extension is extracted
in a manifestly supersymmetric setting~\cite{CS,KScc}. The quantum
modification of the central charge takes place in two ways: 
The short-wavelength quantum fluctuations modify the operator structure of
the central charge, yielding an anomaly, while the long-wavelength
fluctuations affect its expectation value.  
An efficient method is developed to determine the short-wavelength
anomaly  by considering dilatation in superspace.

In Sec.~II we review the derivation of the trace anomaly, 
which is generalized to treat superconformal anomalies
in the succeeding sections.
In Sec.~III we start with a classical BPS soliton in a 2d theory 
with N=1 supersymmetry and study the quantum features of  
domain walls evolving from it in higher dimensions. 
In Sec.~IV we discuss vortices in 3d and 4d theories. 
Section~V is devoted to a summary and discussion.

\section{Dilatation anomaly}

Let us first review a derivation of the trace anomaly~\cite{ACD,CDJ}.
Consider a scalar field $\phi(x)$ (of dimension $d_{\phi}$ in units of
mass) with the action $S[\phi]$ and let $T^{\mu\nu}$ denote a conserved
symmetric energy-momentum tensor.  The current associated with an
infinitesimal scale transformation
$\delta_{\rm D} \phi = (x^{\mu}\partial_{\mu} + d_{\phi}) \phi$ 
in general is written as $x_{\nu}T^{\mu\nu}$ with the conservation law
\begin{equation}
\partial_{\mu}(x_{\nu}T^{\mu\nu}) = \triangle  + j \delta_{\rm D} \phi,
\label{dilatationcurrent}
\end{equation}
where $j$ denotes a source for $\phi$; $\triangle$ stands
collectively for scale breaking, both classical and quantum.
With translational invariance assumed, Eq.~(\ref{dilatationcurrent})
gives rise to the trace identity 
$T^{\mu}_{\mu} = \triangle + j\, d_{\phi}\, \phi$. 

Equation~(\ref{dilatationcurrent}) or
$\int d^{n}x\, (\triangle + j \delta_{\rm D} \phi) =0$ with $n$ being the
dimension of spacetime yields, e.g., for a two-point Green function 
$\langle \phi\phi\rangle$ [with Fourier transform $G(p)$], 
a Ward-Takahashi identity of the form~\cite{CSym,CJ}
\begin{equation}
i G^{\triangle}(p)
= (p^{\nu}\partial/\partial p^{\nu} +n -2d_{\phi}) G(p),
\label{Gtriangle}
\end{equation}
where $G^{\triangle}(p) \stackrel{\rm F.T.}{=}  \langle \phi \phi 
\int d^{n}x\, \triangle \rangle$, i.e., $G(p)$ 
with a zero-momentum insertion of $\triangle$.

In calculating $G(p)$ one encounters ultraviolet (UV) divergences.
The (bare) theory is thus written in terms of field $\phi(x)$, 
a bare mass $m$, some coupling constants $\lambda$ 
(of dimension $d_{\lambda}$) and the UV cutoff $\Lambda$ (of dimension one). 
We go through renormalization as usual and denote renormalized quantities 
with the suffix ${\rm r}$. 
A dimensional analysis then allows one to rewrite Eq.~(\ref{Gtriangle}) 
as
\begin{eqnarray}
i G_{\rm r}^{\triangle}(p)
&=& -{\cal D} G_{\rm r}(p; m_{\rm r}, \lambda_{\rm r}),
\label{dGr} \\
{\cal D} &=& m_{\rm r}\partial/\partial m_{\rm r} 
+ d_{\lambda} \lambda_{\rm r}  \partial/\partial \lambda_{\rm r}.
\label{operatorD}
\end{eqnarray}
(In case a new scale $\mu$ is introduced for
renormalization,  one may simply include $\mu \partial/\partial \mu$ in
${\cal D}$.)   
One can now read off the scale breaking $\triangle$ 
from the renormalized Green function
$\langle \phi_{\rm r} \phi_{\rm r}\rangle$.
Indeed, 
the action of the derivative ${\cal D}$ on $G_{\rm r}$ is essentially 
the same as an insertion into $G_{\rm r}$  of the operator ${\cal D}S$
with the action $S$ now regarded as a function(al) of 
$\phi_{\rm r},m_{\rm r}, \lambda_{\rm r}$ and the cutoff $\Lambda$,
\begin{equation}
\int d^{n}x\, \triangle 
= - {\cal D} S[\phi_{\rm r} ; m_{\rm r},\lambda_{\rm r},
\Lambda].
\label{scalebreakingDS}
\end{equation}

In early approaches~\cite{ACD, CDJ} it was customary 
to rewrite  ${\cal D}G_{\rm r}$ in favor of 
the Callan-Symanzik equations~\cite{CSym} or 
the renormalization group equations so that quantum scale breaking
comes with the associated anomalous dimensions;
we comment on this point later.
Our identification of $\triangle$ by ${\cal D} S$ is simpler in form 
and leads to expressions suited for our analysis of the central
charge anomalies.

It is enlightening to rewrite Eq.~(\ref{scalebreakingDS}) further.   
Consider how $S[\phi_{\rm r} ; m_{\rm r},\lambda_{\rm r}, \Lambda]$
itself responds to dilatation $\delta_{\rm D} \phi_{\rm r}$.
The response is precisely what one would get classically,
$\delta_{\rm D}^{\rm cl}S = - {\cal D}S 
- \Lambda \partial S/\partial \Lambda$, 
which now also detects the presence of the cutoff $\Lambda$.
The true quantum response $- {\cal D}S =\int d^{n}x\, \triangle$
thus equals $\delta_{\rm D}^{\rm cl}S + \Lambda \partial S/\partial
\Lambda \equiv \delta_{\rm D}S$.
(Note in this connection that the renormalized Green function
$G_{\rm r}(p; m_{\rm r}, \lambda_{\rm r})$ has no reference to 
the cutoff $\Lambda$.)
This allows us to write Eq.~(\ref{scalebreakingDS}) as
\begin{equation}
\delta_{\rm D}S = \int d^{n}x\, \triangle 
= \int d^{n}x\, (\triangle^{\rm cl} +  \triangle^{\rm anom})
\label{deltaDS}
\end{equation}
with the quantum breaking (anomaly) given by
\begin{equation}
\int d^{n}x\, \triangle^{\rm anom} = \Lambda (\partial
/\partial \Lambda) S[\phi_{\rm r} ; m_{\rm r},\lambda_{\rm r}, \Lambda].
\label{quantumbreaking}
\end{equation}
This is the basic formula we shall use below. 
We remark that Eq.~(\ref{deltaDS}) is an operator analog of the response
of the effective action $\Gamma [\phi_{\rm c}]$ to dilatation 
$\delta_{\rm D}\phi_{\rm c}= (x^{\mu}\partial_{\mu} + d_{\phi}) \phi_{\rm c}$ 
[of the classical field $\phi_{\rm c} =\langle \phi \rangle_{j}$],
\begin{eqnarray}
\delta_{\rm D} \Gamma [\phi_{\rm c}] 
=  \int d^{n}x\, \langle \triangle \rangle,
\label{deleffectiveaction}
\end{eqnarray}
which follows from Eq.~(\ref{dilatationcurrent}) 
[with  $j = -\delta \Gamma/\delta \phi_{\rm c}$].

\section{Solitons and domain walls}

In this section we study the effect of quantum fluctuations on
solitons and domain walls. 

\subsection{Solitons in two dimensions} 
Let us begin by reviewing some basic features of
supersymmetric theories with solitons in two dimensions.
Consider the Wess-Zumino model~\cite{WB,SVV} described by 
the superfield action
\begin{equation}
 S[\Phi] = \int dz\, \left\{ {1\over{4}}\,
(\bar{D}_{\alpha}\Phi)\, D_{\alpha}\Phi + W(\Phi)
\right\},
\label{sfaction}
\end{equation}
with a real superfield 
$\Phi (z) = \phi (x) + \bar{\theta}\psi (x)  + {1\over{2}}\,\bar{\theta}\theta\, F(x)$,
consisting of a real scalar field $\phi$ and a Majorana spinor field
$\psi_{\alpha} =(\psi_{1},\psi_{2})$, along with an auxiliary field $F$.
Here $z=(x^{\mu},\theta_{\alpha})$ denotes 
the spacetime coordinates $x^{\mu}$ and two Majorana coordinates
$\theta_{\alpha}=(\theta_{1},\theta_{2})$ in N=1 superspace; 
$dz\equiv d^{2}xd^{2}\theta$ with $\int d^{2}\theta\,
{1\over{2}}\,\bar{\theta}\theta =1$; 
$\bar{\theta}\equiv \theta\gamma^{0}  = i(\theta_{2}, -\theta_{1})$ and  
$\bar{\theta}\theta=\bar{\theta}_{\alpha}\theta_{\alpha}=
-2i\theta_{1}\theta_{2}$ with the Dirac matrices $\gamma^{0}=\sigma_{2}$ 
and $\gamma^{1}=i\sigma_{3}$.
The spinor derivatives 
$D_{\alpha}= \partial/\partial \bar{\theta}_{\alpha}
- (\slash\!\!\! p\, \theta)_{\alpha}$ 
and $\bar{D}_{\alpha}\equiv (D\gamma^{0})_{\alpha}$, 
with
$\slash\!\!\! p \equiv \gamma^{\mu}p_{\mu}$ and
$p_{\mu}=i\partial_{\mu}$, 
obey 
$ \{D_{\alpha}, D_{\beta}\}  = 2\, (\slash\!\!\! p \gamma^{0})_{\alpha\beta}$.

In components the action reads $S = \int d^{2}x\,  {\cal L}$  with 
\begin{equation}
 {\cal L} =  {1\over{2}} \bar{\psi} 
\{i {\slash\!\!\!\partial} -  W''(\phi) \} \psi 
+ {1\over{2}}\, (\partial_{\mu}\phi)^{2}\! 
-{1\over{2}}\, [W'(\phi)]^{2},
\label{modelcomponent}
\end{equation}
on eliminating the auxiliary field $F$, via $\delta S/\delta F 
= F + W'(\Phi)=0$; $W'(\phi)=dW(\phi)/d\phi$, etc.
With a superpotential
\begin{equation}
W(\phi)= {m^{2}\over{4\lambda}}\, \phi 
- {\lambda\over{3}}\, \phi^{3}
\label{Wphi}
\end{equation}
having two extrema $W'(\phi)=0$, the model supports a classical 
soliton (kink)  solution 
\begin{equation}
\phi_{\rm sol} (x) =v\tanh(mx^{1}/2),
\label{classicalsoliton}
\end{equation}
governed by the the Bogomol'nyi equation~\cite{B}
\begin{eqnarray}
\partial_{1}\phi = -F = W'(\phi),
\end{eqnarray}
with $v=m/(2\lambda)$. The kink interpolates between the two distinct
vacua  with $\langle \phi \rangle_{\rm vac}=\pm v$ at
spatial infinities $x^{1}=\pm \infty$, and has energy 
$m^{\rm cl}_{\rm sol} = m^{3}/(6\lambda^{2})$.

The action $S$ is invariant under supertranslations 
\begin{equation}
x^{\mu} \rightarrow x^{\mu} + i\bar{\xi} \gamma^{\mu} \theta ,\ \ 
\theta_{\alpha} \rightarrow \theta_{\alpha} +\xi_{\alpha},
\label{suptranslation}
\end{equation}
generated by the supercharge
$\bar{\xi}_{\alpha}Q_{\alpha}$ acting on $\Phi(z)$.
The component fields thereby undergo the supersymmetry transformations
$\delta \phi = \bar{\xi}\psi, 
\delta F = -i \bar{\xi}\gamma^{\mu} \partial_{\mu}\psi$, etc.
In the presence of solitons the supercharge algebra
acquires a central extension~\cite{WO},
\begin{equation}
\{Q_{\alpha}, \bar{Q}_{\beta}\} 
= 2(\gamma_{\lambda})_{\alpha\beta}\, P^{\lambda}
 + 2i(\gamma_{5})_{\alpha\beta}\,Z
\label{scalgebra}
\end{equation}
with the central charge $Z=\int d x^{1}\, \zeta^{0}$
coming from the topological current 
\begin{eqnarray}
\zeta^{\mu} &=& - \epsilon^{\mu\nu} F\partial_{\nu}\phi
\approx \epsilon^{\mu\nu}\partial_{\nu}W(\phi),
\label{topolcurrent}
\end{eqnarray}
where $\gamma_{5}= \gamma^{0}\gamma^{1}$ and $\epsilon^{01} = 1$.

An important fact is that the Bogomol'nyi equation is invariant,
$\delta(\partial_{1}\phi + F) =0$, under supertranslations 
with $\xi_{1}$ alone, 
generated by $\bar{\xi}_{2}Q_{2}\sim \xi_{1}Q_{2}$.
For $Q_{2}$ Eq.~(\ref{scalgebra}) reads $(Q_{2})^{2} = P^{0} -
P^{1} -Z$. The classical soliton therefore is BPS saturated,
\begin{equation}
Q_{2} |{\rm sol}\rangle  = (P^{0} - Z)|{\rm sol}\rangle =0,
\label{BPSsaturation}
\end{equation}
and preserves half of the original supersymmetry.
As a result, via multiplet shortening~\cite{WO,LSV}, the soliton 
is expected to be BPS-saturated at the quantum
level (as long as the supercharge algebra is not afflicted with a
quantum anomaly) and the soliton spectrum
$\langle {\rm sol}|P^{0} |{\rm sol}\rangle =\langle {\rm sol}| Z 
|{\rm sol}\rangle$ should equal the central charge.

The canonical or classical supercharge algebra~(\ref{scalgebra}) 
is a consequence of super-Poincare symmetry.
It implies that the supersymmetry current 
$J^{\mu}_{\alpha}$, topological current $\zeta^{\mu}$ and 
energy-momentum tensor $T^{\mu \lambda}$ form a supermultiplet.  
This structure is best visualized if one considers a superfield
supercurrent~\cite{FZ}, which in the present case is 
a real spinor-vector superfield~\cite{KScc}
\begin{equation}
{\cal V}^{\mu}_{\alpha} 
= -i(D_{\alpha}\bar{D}_{\lambda}\Phi)\, 
(\gamma^{\mu})_{\lambda\beta}\,D_{\beta}\Phi,
\label{sfcurrent}
\end{equation}
with the expansion 
\begin{equation}
{\cal V}^{\mu}_{\alpha}
= J^{\mu}_{\alpha} -2i(\gamma_{\lambda}\theta)_{\alpha} 
T^{\mu \lambda}
+2(\gamma_{5}\theta)_{\alpha} \zeta^{\mu}+\cdots.
\label{sfcurrentcomponent}
\end{equation}
This supercurrent, when properly regulated, is conserved 
at the quantum level~\cite{KScc}, 
$\partial_{\mu}{\cal V}^{\mu}_{\alpha} = 0$, and gives rise to 
a conserved-charge superfield
\begin{equation}
\int dx^{1}{\cal V}^{0}_{\alpha}
= Q_{\alpha} -2i(\gamma_{\mu}\theta)_{\alpha} P^{\mu}
+2(\gamma_{5}\theta)_{\alpha} Z +\cdots,
\label{chargesf}
\end{equation}
which, upon supertranslations~(\ref{suptranslation}),
leads to Eq.~(\ref{scalgebra}).

The supercharge algebra~(\ref{scalgebra}) thus retains the same form in
quantum theory. 
One must, however, note that the central charge, as well as other charges,
may acquire a quantum modification.
Such a quantum anomaly comes from short distances  
and arises as part of the superconformal anomaly,
especially in a manifestly supersymmetric setting.
Actually, the superfield combination
\begin{equation}
 i(\gamma_{\mu}{\cal V}^{\mu})_{\alpha}
=  i(\gamma_{\mu}J^{\mu})_{\alpha}
+2\,\theta_{\alpha} T^{\mu}_{\mu}
+ 2i(\gamma_{\mu}\theta)_{\alpha}\, \epsilon^{\mu \nu}
\zeta_{\nu} +\cdots,
\end{equation}
shows that the topological current
$\zeta^{\nu}$ and the trace  $T^{\mu}_{\mu}$ lie in a multiplet,
i.e., the central charge and dilatation    are related by supersymmetry.
This structure suggests us a general method to determine 
the central charge: One may consider the trace anomaly 
in superspace and  translate it into the superconformal anomaly.

In the present case, naive use of field equations yields a (classical) 
supersymmetric trace identity of the form 
\begin{eqnarray}
 i(\gamma_{\mu}{\cal V}^{\mu})_{\alpha}
&\stackrel{\rm cl}{=}& -2D_{\alpha} W(\Phi).
\label{STItwocl}
\end{eqnarray}
To promote this to quantum theory one may first study renormalization. 
In the 2d Wess-Zumino model only the bare mass $m$ requires infinite
renormalization~\cite{KScc}
\begin{eqnarray}
&&m^{2} = m_{\rm r}^{2} + {\lambda^{2}\over{\pi}} \log
(\Lambda^{2}/m^{2}_{\rm r})
\label{massrenormWZ}
\end{eqnarray}
at the one-loop level, with a UV cutoff $\Lambda$.
Formula~(\ref{deltaDS}) then tells us that this mass renormalization 
leads to a trace anomaly 
$\Lambda \partial W/\partial \Lambda = (\lambda/2\pi)\, \Phi 
= - W''(\Phi)/(4\pi)$. 
See the appendix for an evaluation of the classical breaking 
$\delta_{\rm D}^{\rm cl}S$. 
The full scale breaking is given by 
$\int d^{2}x\,\triangle = - \int dz\, W_{\rm eff}(\Phi)$ with
\begin{equation}
 W_{\rm eff}(\Phi)= W(\Phi)  + {1\over{4\pi}} W''(\Phi) .
\end{equation}
   [Actually a dimensional analysis such that  
\begin{equation}
-  {\cal D} W(\Phi) = 
\Big\{ -d_{W} + \Lambda{\partial\over{\partial \Lambda}}
+d_{\Phi} \Phi_{\rm r} {\partial \over{\partial \Phi_{\rm r}}}
 \Big\} W(\Phi) 
\label{scalebrkg}
\end{equation}
for general $W(\Phi)$ of dimension $d_{W}$ offers a simpler way to reach this
result (on setting $d_{W}=1$ and $d_{\Phi}=0$).]
This breaking $\triangle$ now dictates that Eq.~(\ref{STItwocl}) should read 
\begin{equation}
i(\gamma_{\mu}{\cal V}^{\mu})_{\alpha}  =-2D_{\alpha}
W_{\rm eff}(\Phi)
\end{equation}
in quantum theory, giving the modified central charge,
\begin{equation}
Z =   \int dx^{1}\partial_{1}  W_{\rm eff}(\phi)
= 2\, W_{\rm eff}(\phi)|_{x^{1}=\infty}.
\label{Zanom}
\end{equation}

Now the exact form of the central charge {\em operator} $Z$ is known, but
not the central charge yet. In quantum theory the central charge is given
by  the ground-state expectation value $\langle {\rm sol}| Z|{\rm sol} \rangle 
\equiv \langle  Z \rangle = 2\langle W_{\rm eff}(\Phi)\rangle|_{x^{1}=\infty}$,  
which is only approximately known in powers of $\hbar$.

Let us discuss this point in some detail. To determine the central charge
$ \langle  Z \rangle$ one first has to calculate the effective action 
$\Gamma [\Phi_{\rm c}]$ and learn how the quantum effect modifies 
the superpotential and its extrema. 
Indeed, a one-loop calculation yields a superpotential of the form 
\begin{equation}
{\cal P}_{\rm eff}(\Phi_{\rm c}) =  W_{\rm c} + {1\over{8\pi}}\, 
W''_{\rm c} \Big\{ \ln {\Lambda^{2}\over{(W''_{\rm c})^{2}}} + 2 \Big\},
\label{Peff}
\end{equation}
where $W_{\rm c}\equiv W(\Phi_{\rm c})$ and 
$W''_{\rm c}= -2 \lambda \Phi_{\rm c}$;
the cutoff $\Lambda^{2}$ is eliminated by mass
renormalization~(\ref{massrenormWZ}), the effect of which is to set
$m^{2} \rightarrow m_{\rm r}^{2}$ and 
$\Lambda^{2} \rightarrow m_{\rm r}^{2}$ 
in ${\cal P}_{\rm eff}(\Phi_{\rm c})$.
Actually the divergent piece is part of the one-loop correction to 
the expectation value $\langle W(\Phi) \rangle$
and one can write ${\cal P}_{\rm eff}(\Phi_{\rm c}) = \langle W(\Phi)
\rangle + W''_{\rm c}/(4\pi) = \langle W_{\rm eff}(\Phi)\rangle$. 
One can verify the anomaly $W''(\Phi)/(4\pi)$ by a direct calculation of  
the response $\delta_{\rm D} {\cal P}_{\rm eff}(\Phi_{\rm c})  
= - \langle W_{\rm eff}(\Phi)\rangle + {\rm total\ div.}$ 
Minimizing  ${\cal P}_{\rm eff}(\Phi_{\rm c})$  with respect to $\Phi_{\rm c}$ 
yields the central charge and 
hence the soliton spectrum to $O(\hbar^{0})$~\cite{NSNR,SVV},
\begin{equation}
m^{\rm sol}  = \langle Z \rangle
= m_{\rm r}^{3}/ (6\lambda^{2}) -m_{\rm r}/ (2\pi).
\label{msoliton}
\end{equation}
Here the central-charge anomaly $\sim W''/(4\pi)$  directly fixes the
one-loop correction $-m_{\rm r}/ (2\pi)$ to $m^{\rm sol}$.  
Note that the spectrum is approximate; it is further
modified by higher-loop corrections  [to ${\cal P}_{\rm eff}(v)$]
of long-wavelength nature
$\propto m_{\rm r} (\lambda^{2}/m_{\rm r}^{2})^{\ell -1} \sim
O(\hbar^{\ell -1})$ with $\ell$ loops.

The quantum characteristics of solitons are determined 
by solving the quantum equation of motion 
$\delta \Gamma [\Phi_{\rm c}]/\delta \Phi_{\rm c} =0$.
It is difficult to calculate the corrected soliton profile
explicitly.
Fortunately the soliton spectrum is calculable from 
the asymptotic $(x^{1} \rightarrow \pm \infty)$ profile alone, 
owing to BPS saturation.
A look into an asymptotic effective action  of the form
\begin{equation}
\Gamma [\Phi_{\rm c}] =\int dz\, [
f(\Phi_{\rm c})(\bar{D}_{\alpha}\Phi_{\rm c})(D_{\alpha}\Phi_{\rm c})
+ {\cal P}_{\rm eff}(\Phi_{\rm c}) ]
\end{equation}
reveals that the central charge is given by 
$\langle Z \rangle = 2{\cal P}_{\rm eff}(v)$ with 
$v =\Phi_{\rm c}|_{x^{1}\rightarrow \infty}$ fixed 
by minimizing  ${\cal P}_{\rm eff}(\Phi_{\rm c}) \approx 
\langle W_{\rm eff}(\Phi)\rangle$.
(The Bogomol'nyi equation is best written in terms of 
$\tilde{\Phi}[\Phi_{\rm c}]$ defined by $d\tilde{\Phi}
= \sqrt{f(\Phi_{\rm c})}\,  d\Phi_{\rm c}$.)

\subsection{Domain walls in three dimensions} 

The classical soliton~(\ref{classicalsoliton}) turns into a domain wall 
if one goes to 1+2 dimensions with the same model~\cite{RNW}.
There one again encounters infinite renormalization only for the bare
mass $m$ up to two loops.
Indeed, a direct one-loop calculation yields the effective
superpotential~\cite{KScc}
\begin{equation}
U_{\rm eff}(\Phi_{\rm c}) =  W_{\rm c} + c\, W''_{\rm c}
-{1\over{16\pi}}\, W''_{\rm c}|W''_{\rm c}| ,
\label{Ueff}
\end{equation}
with a linear divergence 
$c \equiv \Lambda/(8\pi \sqrt{\pi})$, which is taken care of by
mass renormalization
\begin{eqnarray}
m^{2} &=& m_{\rm r}^{2} + \delta m^{2} ,\ \ 
\delta m^{2} = 8\lambda^{2} c.
\end{eqnarray}
Minimizing this $U_{\rm eff}(\Phi_{\rm c})$
yields the central charge and hence the domain-wall surface tension to
$O(\hbar^{0})$~\cite{RNW},
\begin{equation}
M^{\rm DW}/L_{2} 
=2U_{\rm eff}(v)={m_{\rm r}^{3}\over{6\lambda^{2}}}
-{m^{2}_{\rm r}\over{8\pi}},
\label{dwtension}
\end{equation}
where $L_{2}$ denotes the total length  in the $x^{2}$ direction.

The supercurrent ${\cal V}^{\mu}_{\alpha}$ in Eq.~(\ref{sfcurrent}),
on the other hand, is adapted to three dimensions in an obvious fashion. 
Using Eq.~(\ref{scalebrkg}), 
with $d_{\Phi} = 1/2$ and $d_{W} = 2$ now, 
allows one to write down the quantum supertrace identity
\begin{eqnarray}
i(\gamma_{\mu}{\cal V}^{\mu})_{\alpha} 
&=&-4D_{\alpha} W_{\rm eff}^{\rm DW}, \nonumber\\
W_{\rm eff}^{\rm DW} &=&  W(\Phi) - {1\over{4}}\,  \Phi  W'(\Phi)
 + \gamma\, \Phi,\
\label{straceDW}
\end{eqnarray}
with $\gamma \equiv - {1\over{8\lambda}}\,   \Lambda
{\partial \over{\partial \Lambda}}
m^{2}(\lambda, m_{\rm r}, \Lambda) 
= - \lambda\, c + O(\hbar^{3/2})$.
This generalizes an earlier result~\cite{KScc} to all orders in $\hbar$.  
We remark that via dilatation one has naturally been led to 
the supertrace identity for the improved supercurrent 
$({\cal V}_{\rm imp})^{\mu}_{\alpha}
= {\cal V}^{\mu}_{\alpha} - (i/4)\, \epsilon^{\mu\nu\rho}
\partial_{\nu}(\gamma_{\rho}D\Phi^{2})_{\alpha}$ rather than the
canonical one.  The $({\cal V}_{\rm imp})^{\mu}_{\alpha}$ 
and ${\cal V}^{\mu}_{\alpha}$ are physically equivalent,
leading the same central charge $\langle Z \rangle$.

In three dimensions the short-distance anomaly 
$\propto \gamma \Phi$ simply works to renormalize $m$ and 
the quantum correction $\sim - m_{\rm r}^{2}/(8\pi)$
to the central charge essentially comes from the expectation value 
$\langle W_{\rm eff}^{\rm DW}(\Phi) \rangle$ as a long-wavelength effect.
The domain-wall tension thus in general receives higher-loop
corrections 
$\propto m_{\rm r}^{2}\, (\lambda^{2}/m_{\rm r})^{\ell -1}$.

\subsection{Domain walls in the 4d Wess-Zumino model}

The classical soliton~(\ref{classicalsoliton}), when embedded in
four dimensions, turns into a domain wall, extending uniformly in the
$x^{2}x^{3}$ plane.
Such a wall again remains BPS-saturated~\cite{CS}.
The 4d model consists of a charged boson and a charged (Weyl)
fermion $\psi_{\alpha}(x)$, assembled in a chiral superfield 
$\Phi(z)=\phi (y) +
\sqrt{2}\, \theta^{\alpha}\psi_{\alpha}(y) + \theta^{2} F(y)$
(with $y^{\mu} \equiv x^{\mu} - i\theta \sigma^{\mu}\bar{\theta}$),
\begin{equation}
S =  \int\! d^{8}z\, \bar{\Phi}\Phi 
+\int\! d^{6}z\, W(\Phi) +\int\! d^{6}\bar{z}\, \bar{W}(\bar{\Phi}),
\end{equation}
where $\bar{\Phi}(z) = \Phi(z)^{\dag}$; 
$z= (x^{\mu},\theta^{\alpha},\bar{\theta}_{\dot{\alpha}})$,
$d^{8}z = d^{4}x\, d^{2}\theta\, d^{2}\bar{\theta}$,
$d^{6}z = d^{4}x\, d^{2}\theta$, etc. 
The holomorphic superpotential $W(\Phi)$ takes the same form as
that in Eq.~(\ref{Wphi}) and  $\bar{W}(\bar{\Phi}) = [W(\Phi)]^{\dag}$.
[We adopt superspace notation of Ref.[\onlinecite{WB}] but 
with metric $(+ - - -)$; 
$D_{\alpha}= \partial/\partial \theta^{\alpha} -
(\sigma^{\mu}\bar{\theta})_{\alpha} p_{\mu}$ and
$\bar{D}^{\dot{\alpha}}  = \partial/\partial\bar{\theta}_{\dot{\alpha}} 
- (\bar{\sigma}^{\mu}\theta)^{\dot{\alpha}}p_{\mu}$.]
The Bogomol'nyi equation $\partial_{1}\phi +F =0$, governing the 
classical solution~(\ref{classicalsoliton}), is invariant  under
supertranslations 
\begin{equation}
\delta \theta = \xi,\
\delta\bar{\theta} =\bar{\xi},\ 
\delta x^{\mu} =i\xi \sigma^{\mu}\bar{\theta}
- i\theta\sigma^{\mu}\bar{\xi},
\label{supertransl}
\end{equation}
with $\xi_{\alpha}$ constrained so that $\xi = -i\sigma^{1}\bar{\xi}$, 
and the domain wall preserves half of  4d N=1 supersymmetry.

The 4d Wess-Zumino model requires only infinite wave-function
renormalization and  we set 
$\Phi = \sqrt{Z_{\Phi}} \Phi_{\rm r}$.  
The chiral superpotential $W(\Phi)$ receives no quantum correction 
in perturbation theory~\cite{GSR,Seib} and one may set 
$m_{\rm r} = m\, Z_{\Phi}$ and $\lambda_{\rm r} 
=\lambda\, Z_{\Phi}^{3/2}$; its extrema $W'(\Phi)=0$ arise at $\Phi_{\rm
r}= \pm v_{\rm r}$ with 
$v_{\rm r} \equiv m_{\rm r}/(2\lambda_{\rm r})$.

The domain-wall spectrum, being BPS-saturated, is
calculated from a long-wavelength effective action of the form
\begin{equation}
\Gamma[\Phi_{\rm c}] =  \int\! d^{8}z\, f(M_{\rm r})
+ \left\{ \int\! d^{6}z\, W(\Phi_{\rm c}) + {\rm h.c.}\right\},
\label{WZeffaction}
\end{equation}
where $M_{\rm r} = (\bar{\Phi}_{\rm c})_{\rm r} (\Phi_{\rm c})_{\rm r}$.
Actually, a direct one-loop calculation yields
\begin{equation}
f(M_{\rm r}) \approx M_{\rm r} \big\{ 1 - \alpha\, 
\big[ \ln (M_{\rm r}/v_{\rm r}^{2}) -1\big] \big\}
\label{oneloopWZ}
\end{equation}
with the choice $Z_{\Phi} = 1- \alpha\, \ln (\Lambda^{2}/m_{\rm r}^{2})$;
$\alpha \equiv \lambda_{\rm r}^{2}/(8\pi^{2})$. 
One may use $\Phi'_{\rm c} = \Phi_{\rm c} 
\{1 - \alpha [\ln (\Phi_{\rm c}/v) - 1/2] \}$ 
in completing Bogomol'nyi squares. 
One then learns that extremizing $W(\Phi_{\rm c})$ yields 
the central charge and that the domain-wall tension 
retains the classical value
\begin{equation}
\langle P^{0}\rangle/\Omega = 4W(v)
=  m_{\rm r}^{3}/(3\lambda_{\rm r}^{2})
=  m^{3}/(3\lambda^{2})
\label{spectrumWZ}
\end{equation}
with $v = m/(2\lambda)$; $\Omega \equiv \int dx^{2}dx^{3}$.

We shall now examine this conclusion in terms of the supertrace
identity. To this end it will be useful to first review some general 
properties of the supercurrent in  4d theories~\cite{FZ,CPS}. 
For 4d N=1 theories the supercurrent is given
by a real vector superfield ${\cal R}_{\alpha\dot{\alpha}}(z) 
=(\sigma_{\mu})_{\alpha\dot{\alpha}}{\cal R}^{\mu}(z) $,
with the expansion
\begin{equation}
{\cal R}^{\mu}(z) 
= R^{\mu}  -i \theta^{\alpha} j_{\alpha}^{\mu}
+ i\bar{\theta}_{\dot{\alpha}}(\bar{j}^{\mu})^{\dot{\alpha}} 
-2 \theta\sigma_{\lambda}\bar{\theta}\, v^{\mu\lambda}
+ \cdots,  
\end{equation}
where $R^{\mu}$ is the R-current,  $(j^{\mu}, \bar{j}^{\mu})$
are related to the supersymmetry currents and $ v^{\mu\lambda}$
to the energy-momentum tensor.  The associated supersymmetric trace
identity, in general, takes the form  
\begin{eqnarray}
\bar{D}^{\dot{\alpha}}{\cal R}_{\alpha\dot{\alpha}}=
{\sigma}^{\mu}\bar{D}{\cal R}_{\mu}= D_{\alpha}{\cal C} +
\bar{D}^{2}D_{\alpha}\Xi,
\label{DbarR}
\end{eqnarray}
which leads to the conservation law
\begin{eqnarray}
&&\partial_{\mu}{\cal R}^{\mu} 
=- i \textstyle{1\over{4}}(D^{2}{\cal C} - \bar{D}^{2}\bar{\cal C} ).
\end{eqnarray}
The form of the  superconformal breaking
is constrained owing to boson-fermion balancing.
The breaking may either form  a chiral multiplet of
the form $D_{\alpha}{\cal C}$ with a chiral superfield ${\cal C}$ or 
belong to a linear multiplet of the form
$\bar{D}^{2}D_{\alpha}\Xi$ with a real superfield $\Xi$.
The $\Xi$ may be traded for ${\cal C}$ to some extent 
by a redefinition of the supercurrent, but this is not always the case, 
as we shall see later.  Actually either a chiral or linear supermultiplet 
is allowed for the breaking, but not both at the same time~\cite{CPS}, 
in order for ${\cal R}^{\mu}(z)$ to contain conserved 
supersymmetry currents $(J_{\alpha}^{\mu}, \bar{J}^{\mu\dot{\alpha}})$ 
and energy-momentum tensor $T^{\mu\lambda}$.
(i) For $\Xi=0$  one can write $J_{\alpha}^{\mu}
=j^{\mu}_{\alpha}-(\sigma^{\mu}\bar{\sigma}^{\nu} j_{\nu})_{\alpha}$
and
$T^{\mu\lambda} = v^{\mu\lambda} - g^{\mu\lambda}v^{\rho}_{\rho}$
with $T^{\mu}_{\mu} = -3v^{\rho}_{\rho} = - (3/2) ({\cal C}|_{\theta^{2}} 
+ \bar{\cal C}|_{{\bar{\theta}}^{2}})$.
(ii) For ${\cal C}=0$, instead, one can take
$J_{\alpha}^{\mu} =
j^{\mu}_{\alpha}$ and $T^{\mu\lambda} = v^{\mu\lambda}$
with $T^{\mu}_{\mu} = -{1\over{4}}\,
D^{\alpha}\bar{D}^{2}D_{\alpha}\Xi|_{0}$.
In view of Eq.~(\ref{deltaDS}) this structure is neatly
encoded in a response to dilatation of the form
\begin{eqnarray}
\delta_{\rm D}S  = -{3\over{2}} \Big[ \int d^{6}z\, {\cal C} 
+ {\rm h.\, c.} \Big] - 4 \int d^{8}z\,  \Xi. 
\label{dDGamma}
\end{eqnarray}
Note that Eq.~(\ref{dDGamma}) remains invariant under a possible
redefinition of the supercurrent 
\begin{eqnarray}
{\cal R}'_{\alpha\dot{\alpha}} = 
{\cal R}_{\alpha\dot{\alpha}} + [D_{\alpha}, \bar{D}_{\dot{\alpha}}]X,
\label{currentredef}
\end{eqnarray} 
${\cal C}' = {\cal C} + \bar{D}^{2}X$ and 
$\Xi' = \Xi + 3X$ with real $X$.

Let us now calculate the scale breaking $\int d^{4}x\,\triangle$
for the 4d model.
The potential $W(\Phi)$ leads to classical breaking 
$- \int d^{6}z\,(3 W  - \Phi\, W' )$ while the kinetic term yields 
quantum breaking $-2\gamma_{\Phi} \int d^{8}z\, \bar{\Phi}\Phi$
with
$\gamma_{\Phi} = - {1\over{2}} (\Lambda
\partial/ \partial \Lambda)
\ln Z_{\Phi}(\lambda_{\rm r},m_{\rm r}, \Lambda) 
=\lambda_{\rm r}^{2}/(8\pi^{2}) + O(\hbar^{2})$.
Formula~(\ref{dDGamma}) then allows one to identify 
the superconformal breaking ${\cal C}$ and $\Xi$.
One may take, e.g., $\Xi=0$ and
\begin{equation}
{\cal C} =    2W(\Phi)  - \textstyle{2\over{3}}\, \Phi\, W'(\Phi) 
- {1\over{6}}\, \gamma_{\Phi}\bar{D}^{2}(\bar{\Phi}\Phi)
\label{Cwz} 
\end{equation}
in Eq.~(\ref{DbarR}),
up to a redefinition of the supercurrent.
This confirms an earlier result~\cite{CS}
and, with $\gamma_{\Phi}\approx \lambda_{\rm r}^{2}/(8\pi^{2})$, 
reproduces the one-loop anomaly obtained by direct
calculations~\cite{GW,KSsc}. (The explicit form of ${\cal
R}_{\alpha\dot{\alpha}}$ is read from
Eq.~(\ref{RHiggscl}) with only $\Phi$ retained and $w\rightarrow 2/3$.)

From this supertrace identity one can conclude~\cite{CS} that 
the supercharge algebra
$\{Q_{\alpha}, \bar{Q}_{\dot{\alpha}}\} =
2(\sigma_{\mu})_{\alpha\dot{\alpha}} P^{\mu}$ 
is normal while
$\{Q_{\alpha}, Q_{\beta}\} = -2i(\sigma^{(3)})_{\alpha\beta}\, U$
has a central charge in the presence of a domain wall,
\begin{equation}
U\! =\! \int\! d^{3}x\,
 \partial_{1}\bar{\cal C}|_{0} 
\!=\!  {m_{\rm r}^{2}\over{3\lambda_{\rm r}}} \int\! d^{3}x\,
 \partial_{1}\bar{\phi}_{\rm r},
\end{equation}
which derives from $(i/4)D^{2}{\cal R}_{\mu}|_{0}$ 
$= -(i/4)(\bar{\sigma}_{\mu}D)^{\dot{\alpha}}
 D^{\beta} R_{\beta\dot{\alpha}}$ 
$= \partial_{\mu}\bar{\cal C}|_{0}$.
The resulting central charge $\langle U\rangle$ correctly 
reproduces Eq.~(\ref{spectrumWZ}), with 
$\langle \bar{\phi}_{\rm r} \rangle  \rightarrow  v_{\rm r}$
for $x^{1} \rightarrow \infty$.

Some remarks are in order here.
(i) The superconformal anomaly $\propto 
\gamma_{\Phi}\bar{D}^{2}(\bar{\Phi}\Phi)$  in ${\cal C}$ does not
contribute to the central charge because the expectation value 
$\langle \bar{\Phi}\Phi\rangle$ tends to a constant
in the asymptotic vacuum region $x^{1}\! \rightarrow \pm \infty$ 
(where $\Phi_{\rm c}\! \rightarrow \pm v$
with $F\! \rightarrow\!  0$).
(ii) The central charge $\langle U\rangle$ is uniquely determined 
because $2W - (2/3)\, \Phi\, W' 
= (m_{\rm r}^{2}/3\lambda_{\rm r})\, \Phi_{\rm r}$ in ${\cal C}$ 
is {\it linear} in  $\Phi_{\rm r}$.
One can instead use $2W(\Phi)$ equally well, in accord with
Eq.~(\ref{spectrumWZ}).  This is because, owing to the nonrenormalization
theorem,  the chiral potential $\langle \Phi W'(\Phi) \rangle$ differs
from the classical one $\Phi_{\rm c} W'(\Phi_{\rm c})$ only by terms
$\propto \bar{D}^{2}\langle \cdots \rangle$; thus 
$\langle \Phi W'(\Phi) \rangle \rightarrow 0$ asymptotically for
$W'(\Phi_{\rm c}) \rightarrow 0$. [In the 2d and 3d case, in contrast,
the short-distance anomaly makes
$\langle \Phi W'(\Phi) \rangle \not\rightarrow 0$ asymptotically.]

In the 4d model  the central charge and the domain-wall
spectrum turn out to be stable against quantum corrections.
It is the nonrenormalization of the superpotential~\cite{GSR}, or
holomorphy~\cite{Seib} underlying 4d supersymmetry, 
that is crucial for this stability.

\section{Vortices}

The supersymmetric Higgs model~\cite{Fayet} in 1+2 dimensions 
supports vortices that are BPS-saturated classically. 
Early studies~\cite{Schm,LM} based on index theorems~\cite{EW,LLM} 
found no quantum correction to the vortex spectrum.
In contrast, recent direct calculations~\cite{Vass,RNWvor},
emphasizing the importance of regularization,
found finite corrections to both the central charge and spectrum, with
BPS saturation maintained.  
The origin of such (apparent) corrections was subsequently clarified 
by formulating the problem in superspace~\cite{KSvor}.
In this section we reexamine this problem
by extending our analysis to vortices; 
we also study quantum features of vortices in a 4d model,
discussed  earlier in a classical theory~\cite{GS}.

Let us start with a 4d model with N=1 supersymmetry~\cite{Fayet},
described by the superspace action
\begin{equation}
S = \int\! d^{6}z\, {1\over{2}}\, W^{\alpha} W_{\alpha} 
+ \int\! d^{8}z \left(-2 \kappa V + {\cal M}\right), 
\label{superHiggs}
\end{equation}
with  ${\cal M} =  \bar{\Phi} e^{2eV}\Phi
+\bar{\Phi}_{-} e^{-2eV}\Phi_{-}$ and $W_{\alpha} = -{1\over{4}}\,
\bar{D}^{2}D_{\alpha}V$.  
Here a pair of oppositely-charged matter superfields $(\Phi, \Phi_{-})$
is necessary to avoid a gauge anomaly.
The real superfield $V(z) = \theta \sigma^{\mu}\bar{\theta}\,
a_{\mu} + \theta^{2}\bar{\theta}\bar{\chi}
+ \bar{\theta}^{2}\theta \chi
+ {1\over{2}}\theta^{2}\bar{\theta}^{2} D + \cdots$ contains 
a gauge field $a_{\mu}$ and a gaugino field $\chi_{\alpha}$.

The Fayet-Iliopoulos (FI) term~\cite{FIL} $-2\kappa V$ serves to
introduce spontaneous breaking of U(1) gauge invariance.
Let us take $\kappa > 0$ and $e>0$. 
The bosonic potential has vanishing flat minima for 
$\langle \bar{\phi}\phi - \bar{\phi}_{-}\phi_{-} \rangle_{\rm vac} 
= \kappa/e$, with supersymmetry maintained.

In generic vacuum configurations 
$\langle \phi_{-} \rangle_{\rm vac} \not=0$ 
there arise no regular vortices~\cite{PRTT}. 
We are interested in a special configuration 
$\langle \phi_{-} \rangle_{\rm vac} =0$ and
$\langle \phi\rangle_{\rm vac} \equiv v = \sqrt{\kappa/e}$, 
where the model supports classical vortex solutions governed 
by the Bogomol'nyi equation~\cite{B,GS},
\begin{equation}
(\nabla_{3} \pm i\nabla_{1})\phi =0,\ 
f_{13} \mp D=0, 
\label{Beq}
\end{equation}
with $D= \kappa - e\bar{\phi}\phi$
and $a_{0} = a_{2} =\phi_{-} = 0$;
 $\nabla_{\mu} = \partial_{\mu} +iea_{\mu}$ and 
$f_{\mu\nu} = \partial_{\mu}a_{\nu} - \partial_{\nu}a_{\mu}$;
here the vortex is taken to lie along the $x^{2}$ axis.
We take the upper sign for vortices, 
with the boundary condition
$\phi(x) \rightarrow  v e^{in_{\rm v}\theta}$ and 
$a_{k} \rightarrow -(n_{\rm v}/e)
\partial_{k}\theta$ for $|{\bf x}|\rightarrow \infty$, 
where $\tan \theta =x^{1}/x^{3}$.
A vortex of vorticity $n_{\rm v}=1,2,...$ carries 
flux $\oint dx^{k}a^{k} =2\pi\, n_{\rm v}/e$
and has tension $E_{\rm cl}$  equal to the central charge
\begin{equation}
Z_{\rm cl}/L_{2} =\int d^{2}{\bf x}_{\perp}\,
\epsilon^{ij}\partial_{i}(\kappa a_{j} + i\bar{\phi}D_{j}\phi) =  2\pi
n_{\rm v}\, \kappa/e
\end{equation}
with $\epsilon^{13}=1$.
The same vortices survive in a 3d N=2 model~\cite{Schm,ENS,LM}, 
obtained via dimensional reduction.

The classical vortex solutions, obeying Eq.~(\ref{Beq}), are
invariant under supertranslations~(\ref{supertransl}) 
with $\xi_{\alpha}$ constrained so that
$\xi (\sigma^{3} -i\sigma^{1})=0$ or $\xi_{1} = i\xi_{2}$.
The vortices therefore are BPS-saturated, 
$(Q_{1} -iQ_{2}) |{\rm vor}\rangle = 0$, and preserve half of 
the 4d N=1 (or 3d N=2) supersymmetry.

The supercurrent for the present model is~\cite{CPStwo}
\begin{eqnarray}
{\cal R}_{\alpha\dot{\alpha}} &=& 
- 2 W_{\alpha}\bar{W}_{\dot{\alpha}}
+(D_{\alpha}e^{2eV}\Phi)
e^{-2eV}(\bar{D}_{\dot{\alpha}}e^{2eV}\bar{\Phi}) \nonumber\\
&& -{w\over{2}}\,
[D_{\alpha}, \bar{D}_{\dot{\alpha}}] (\bar{\Phi} e^{2eV}\Phi) + \cdots, 
\label{RHiggscl}
\end{eqnarray}
with the $\Phi_{-}$ sector recovered by substitution 
$\Phi \rightarrow \Phi_{-}$ and $e\rightarrow -e$.
The R-weight $w$ of the matter fields $(\Phi, \Phi_{-})$ is left
arbitrary in the present model owing to a global U(1) symmetry
of the action $S$. [Normally one has to set 
$w= 2/3$ for four dimensions and $w = 1/2$ for three dimensions.]
One can derive this gauge-invariant supercurrent
${\cal R}_{\alpha\dot{\alpha}}$ directly from the
action $S$ by a superspace Noether theorem~\cite{KSsc}.
The classical supertrace identity thereby reads
\begin{eqnarray}
\bar{D}^{\dot{\alpha}}{\cal R}_{\alpha\dot{\alpha}} 
= (\sigma_{\mu}\bar{D})_{\alpha}{\cal R}^{\mu}
\stackrel{\rm cl}{=} 4\kappa W_{\alpha}
+ \bar{D}^{2}D_{\alpha}\Xi_{0}
\label{strclassical}
\end{eqnarray}
with $\Xi_{0} = -{1\over{4}}(3w-2){\cal M}$ vanishing
correctly for $w=2/3$.

Let us now move to quantum theory.
For quantization we adopt the superfield Feynman gauge but 
suppress the gauge-fixing term $S^{\rm gf}$
whose effect will be recovered later.
In the 4d model one has to renormalize the charge and fields, with
rescaling  
$e = e_{\rm r}/\sqrt{Z_{3}}$, $V=\sqrt{Z_{3}}V_{\rm r}$  and 
$(\Phi,\Phi_{-}) =
\sqrt{Z_{2}}\, (\Phi_{\rm r}, \Phi_{-\rm r})$.
Renormalization of the FI term in general takes place only 
at the one-loop level~\cite{FNPRS}.
In the present case, however, the two oppositely-charged fields
$(\Phi, \Phi_{-})$ combine to cancel divergences, 
and one can define the renormalized FI coupling by a simple rescaling
$\kappa_{\rm r} =
\sqrt{Z_{3}}\, \kappa$ so that $\kappa/e =\kappa_{\rm r}/e_{\rm r}$.

To determine the scale breaking $\int d^{4}x\, \triangle$ 
one may use 
${\cal D}=2\kappa_{\rm r} \partial/ \partial \kappa_{\rm r}$ 
together with the formula~(\ref{quantumbreaking}); see also the appendix.
The result is
\begin{equation}
\int d^{8}z\, ( 4\kappa V 
-2\gamma_{2} {\cal M}) \nonumber\\
- \Big\{ \!\int\! d^{6}z\, \textstyle{1\over{2}}\, 
\gamma_{3} W^{\alpha} W_{\alpha} + {\rm h.c.} \Big\},
\label{scbHiggs}
\end{equation}
with 
$\gamma_{2} = - {1\over{2}} (\Lambda \partial/ \partial \Lambda)
\ln Z_{2}(e_{\rm r}, \kappa_{\rm r}, \mu, \Lambda)$ 
and analogously for $\gamma_{3}$ with $Z_{3}$;
$\gamma_{3}= -\gamma_{2} = e_{\rm r}^{2}/(8\pi^{2})$ at one loop.
A comparison with Eq.~(\ref{dDGamma}) then leads to 
a supertrace identity of the form
\begin{equation}
\bar{D}^{\dot{\alpha}}{\cal R}_{\alpha\dot{\alpha}}
= 4 \kappa\, W_{\alpha} + {1\over{2}}\,
\gamma_{2} \bar{D}^{2}D_{\alpha}{\cal M} 
+ {1\over{3}}\,  \gamma_{3}\, D_{\alpha} W^{2}.
\label{DbarRHiggs} 
\end{equation}
(Here  we have set $\Xi_{0} =0$.) 
The form of the anomalies $\propto
\gamma_{2}, \gamma_{3}$ is somewhat arbitrary
and would depend on the regularization method one uses;
in the above we have chosen a form attained by the path-integral
method. 
[See Ref.~\cite{KSvor}.
A direct calculation shows that 
$2 W_{\alpha} \delta S/\delta V$ in Eq.~(3.7) there yields
${1\over{2}}\,\gamma_{2} \bar{D}^{2}D_{\alpha}{\cal M}$ while 
the remaining terms give $(1-w)\, \gamma_{3}\, D_{\alpha} W^{2}$ 
at the one-loop level, though we omit the detail.]

There is a problem with the supertrace identity~(\ref{DbarRHiggs}).
It contains both linear and chiral scale breaking. 
One can transform $\kappa W_{\alpha}$ into a chiral form
$\propto D_{\alpha}\bar{D}^{2}V$ via a current redefinition 
by $\propto [D_{\alpha}, \bar{D}_{\dot{\alpha}}]V$, which would, however,
spoil gauge invariance of the supercurrent.   For $\kappa \not = 0$ it
would thus appear unlikely that one can construct conserved supercharges
and energy-momentum tensor.

We point out that there is a way out if one notes the supersymmetric
chiral anomaly or the Konishi anomaly~\cite{KK,KKKS}, 
which takes the form
\begin{eqnarray}
{1\over{4}}\, \bar{D}^{2}{\cal M} = \gamma_{\rm c} W^{\alpha}W_{\alpha},
\label{susychiral}
\end{eqnarray}
where 
$\gamma_{\rm c} = (2e)^{2}/(32\pi^{2}) 
= e^{2}/(8\pi^{2})$ receives no higher-order corrections.
One can now cast, 
via a gauge-invariant redefinition of the supercurrent
[with $X \propto {\cal M}$ in Eq.~(\ref{currentredef})], 
the supertrace in the form
 \begin{eqnarray}
\bar{D}^{\dot{\alpha}}{\cal R}_{\alpha\dot{\alpha}}
&=& 4 \kappa\, W_{\alpha} 
+ \beta\bar{D}^{2}D_{\alpha}{\cal M}
\label{STplusChiral}
\end{eqnarray}
with $\beta = {1\over{2}}\, \gamma_{2} 
-  {1\over{4}}\, \gamma_{3}/ \gamma_{\rm c}$.
With only breaking of a linear multiplet, 
the supercurrent is now conserved, 
$\partial_{\mu}{\cal R}^{\mu}=0$,
leading to a conserved-charge superfield 
\begin{equation}
\int\! d^{3}{\bf x}\, {\cal R}^{0}
= Q_{\rm R} - i\theta Q + i\bar{\theta}\bar{Q}  
- 2 \theta\sigma_{\mu}\bar{\theta}\, P^{\mu} 
- 2 \theta\sigma_{2}\bar{\theta}\, Z,
\label{sfcharge}
\end{equation}
consisting of the R charge $Q_{\rm R}$, supercharges 
$(Q_{\alpha}, \bar{Q}^{\dot{\alpha}})$, 
four-momentum $P^{\mu}= \int d^{3} {\bf x}\, T^{0 \mu}_{\rm sym}$ 
and central charge $Z=\int d^{3}{\bf x}\, T^{02}_{\rm asym}$, 
with other charges  vanishing.

The central charge comes from the antisymmetric component of
$T^{\mu\lambda}$, which is a total divergence,
\begin{eqnarray}
 T^{\mu\lambda}_{\rm asym} &=& 
 - \epsilon^{\mu\lambda\nu\rho}  \partial_{\nu} \big\{  
 \kappa a_{\rho}  -\beta j^{5}_{\rho}+ \textstyle{1\over{4}}\, R_{\rho} 
 \big\}, 
\label{Tasymfromti} 
\end{eqnarray}
with    
$j^{5}_{\rho}= {1\over{4}}\, (D\sigma_{\rho}\bar{D} 
- \bar{D}\bar{\sigma}_{\rho}D){\cal M}|_{0}$ being
the matter axial current; $\epsilon^{0123}=1$.  
Note that $j^{5}_{\rho}$ and $R_{\rho}$ are gauge-invariant while 
$a_{\rho}$ is not. 
One may now imagine piercing the vacuum state with a
vortex adiabatically, or make a singular gauge transformation.  
Then only the FI coupling detects the asymptotic winding of the vortex
field $\phi(x)\sim v e^{in_{\rm v}\theta}$, yielding the central-charge
operator 
\begin{equation}
Z/L_{2}= \kappa\!  \int\! d^{2}{\bf x}_{\perp}\, f_{13} = \kappa_{\rm r}\!
\oint dx^{k} a_{\rm r}^{k} 
\label{Zquantum}
\end{equation}
and the central charge 
$\langle {\rm vor}| Z |{\rm vor}\rangle/L_{2} 
= 2\pi n_{\rm v}\kappa_{\rm r}/e_{\rm r}$.

 The charge superfield~(\ref{sfcharge}), upon
supertranslations~(\ref{supertransl}), 
gives rise to the supercharge algebra 
\begin{equation}
\{Q_{\alpha}, \bar{Q}_{\dot{\alpha}}\} 
= 2 (\sigma_{\mu})_{\alpha\dot{\alpha}}\, P^{\mu} 
+ 2(\sigma_{2})_{\alpha\dot{\alpha}}\,Z ,
\label{scaHiggs}
\end{equation}
and $\{Q_{\alpha}, Q_{\beta}\} = 0$.
The vortex then remains BPS-saturated, with 
$(Q_{1} - iQ_{2})|{\rm vor} \rangle =0$, and the vortex tension 
\begin{equation}
\langle {\rm vor}| P^{0} |{\rm vor}\rangle /L_{2}  
= 2\pi n_{\rm v}\kappa_{\rm r}/e_{\rm r}= 2\pi n_{\rm v}\kappa/e,
\label{vortexspect}
\end{equation}
survives renormalization.

In contrast, the spectrum $m \sim \sqrt{2} ev$ of the elementary
excitations (of $a_{\mu}, \psi, \cdots$) is corrected at higher loops.
To see this explicitly let us calculate the effective action 
to one loop by retaining only terms with 
no $D_{\alpha}$ and $\bar{D}_{\dot{\alpha}}$ acting on 
$(\Phi_{\rm c}, V_{\rm c})$.
Such a calculation was done earlier for three dimensions, 
and one may now simply evaluate Eq. (A8) of Ref.~\cite{KSvor} 
in four dimensions with 
$\mu^{2} \rightarrow 2e_{\rm r}^{2}{\cal M}_{\rm r}$ there;
${\cal M}_{\rm r} = {\cal M}/Z_{2}$.
The result is
\begin{equation}
\Gamma^{(1)}
=  \int d^{8}z\, 
{e^{2}_{\rm r}\over{8\pi^{2}}}\, {\cal M}_{\rm r}
\Big\{ \ln{2e_{\rm r}^{2} {\cal M}_{\rm r}\over{\mu^{2}}}
- 1 \Big\}
\label{Gammaone}
\end{equation}
with the choice $Z_{2} = 1 + (e^{2}_{\rm r}/8\pi^{2})\,  \ln
(\Lambda^{2}/\mu^{2})$; this gives
$\gamma_{2} = - e^{2}_{\rm r}/(8\pi^{2})$ at one loop. 
 [Note that a direct calculation of the response 
$\delta_{\rm D}\Gamma^{(1)}
=-\int d^{8}z\, 2\gamma_{2} {\cal M}_{\rm r}$ verifies
Eq.~(\ref{scbHiggs}).]
Minimizing the effective action
$\Gamma_{\rm eff}
\approx  \int d^{8}z\, ( {\cal M}_{\rm r} -2\kappa_{\rm r} V_{\rm r})
+ \Gamma^{(1)}$
with respect to $V_{\rm c}$ within the asymptotic (vacuum) region 
then yields
\begin{equation}
\kappa_{\rm r}/e_{\rm r} = v_{\rm r}^{2}  \Big\{ 
1  + (e^{2}_{\rm r}/8\pi^{2})\, \ln (2e_{\rm r}^{2}v_{\rm
r}^{2}/\mu^{2})
\Big\},
\label{krbyer}
\end{equation}
which relates $v_{\rm r} = \langle \Phi_{\rm r} \rangle_{\rm vac}$ 
to the central charge $\propto \kappa_{\rm r}$.
This tells us that $v_{\rm r}$ retains the classical form
$\sqrt{\kappa_{\rm r}/e_{\rm r}}$ to one loop if one chooses 
the scale $\mu = m_{\rm r} \equiv \sqrt{2}\, e_{\rm r} v_{\rm r}$.  
It is, however, clear that Eq.~(\ref{krbyer}) is further
modified at higher loops, so is the elementary-excitation spectrum.

Here some remarks are in order as to the effect of gauge fixing. 
For the superfield Feynman gauge one considers the gauge-fixing action 
$S^{\rm gf} =\int d^{8}z\, 
 (-{1\over{8}})\, (\bar{D}^{2}V_{\rm r})D^{2}V_{\rm r}$.
It gives rise to an extra component~\cite{CPStwo}
${\cal R}_{\alpha\dot{\alpha}}^{\rm gf} = (1/24)
[-(D_{\alpha}\bar{D}^{2}V_{\rm r})\bar{D}_{\dot{\alpha}}D^{2}V_{\rm r} +
\cdots]$ for ${\cal R}_{\alpha\dot{\alpha}}$.
The net effect is to replace ${\cal R}_{\alpha\dot{\alpha}}$ 
by ${\cal R}_{\alpha\dot{\alpha}} 
+ {\cal R}_{\alpha\dot{\alpha}}^{\rm gf}$ in Eq.~(\ref{STplusChiral}) 
and to add explicit breaking $D_{\alpha}\bar{D}^{2}B^{\rm gf}$ with
$B^{\rm gf}=(1/48) V_{\rm r}[D^{2},\bar{D}^{2}]V_{\rm r}$ on the
right-hand side. 
Actually these additional terms  ${\cal R}^{\rm gf}_{\alpha
\dot{\alpha}}$ and $D_{\alpha}\bar{D}^{2}B^{\rm gf}$
are pure contact terms and vanish among physical states~\cite{CPStwo}. 
This fact essentially leaves our foregoing analysis intact. 
Alternatively one can formulate the supertrace identity in a manifestly
(background) gauge-invariant way~\cite{KSvor}.

Let us finally take a brief look into vortices in the 3d Higgs model.
In three dimensions only the FI term requires infinite renormalization.
Accordingly the model~(\ref{superHiggs}) is a finite theory and 
no anomaly arises.  
It is enlightening to consider a somewhat nontrivial case 
where only one matter field $\Phi$ is retained; 
in three dimensions there is no gauge anomaly and one may keep
only $\Phi$. 
The single-$\Phi$ model also supports classical vortices governed by 
Eq.~(\ref{Beq}).
For this model the one-loop correction to the FI term 
$-2\kappa V$ is linearly divergent, and the divergence is removed 
via renormalization~\cite{KSvor}
\begin{equation}
\kappa = \kappa_{\rm r} + \delta \kappa, \  
\delta \kappa = (e/4\pi \sqrt{\pi})\,  \Lambda.
\label{kvsv}
\end{equation}
Formula~(\ref{quantumbreaking})  then implies the
presence of quantum scale breaking
$-2\delta \kappa \int d^{7}z\, V$ (which is divergent). 
Indeed, acting on the action with 
${\cal D}= {3\over{2}}\kappa_{\rm r}\partial/\partial
\kappa_{\rm r} + {1\over{2}}e\partial/\partial e$ yields
\begin{equation}
\delta_{\rm D}S = \int d^{7}z\, \Big[ 
2 (\kappa -\delta \kappa) V +  (\kappa -e{\cal M}) V
\Big].
\label{fullbreaking}
\end{equation}
One can rewrite $(\kappa - e {\cal M}) V$
as $-{1\over{2}}\, V(DW + \delta S/\delta V)$. 
The equation-of-motion term $V(\delta S/\delta V)$, when regulated,
vanishes in three dimensions~\cite{KSvor}, and 
$\int d^{7}z (-{1\over{2}} V DW)$ equals 
$\int d^{5}z\, {1\over{2}}\, W^{2}$.
Interestingly one can get to this result by assigning $d_{V}=0$; 
this corresponds to rescaling $eV \rightarrow V$ in considering
dilatation; see the appendix.

The classical supertrace~(\ref{strclassical}), 
on the other hand, applies to the 3d case as well, 
although one has to note that on reduction
$(\sigma_{2}\bar{D})_{\alpha}{\cal R}^{2}$ turns into superconformal
breaking. The 3d form of the supertrace thereby is written as
\begin{eqnarray}
&&(\sigma_{\mu'}\bar{D})_{\alpha}{\cal R}^{\mu'} 
\stackrel{\rm cl}{=} 2\kappa W_{\alpha} +
\bar{D}^{2}D_{\alpha}\Xi_{0}^{\rm (3d)}, 
\label{threeDstrcl} \\
&&\Xi_{0}^{\rm (3d)} = 
 {\textstyle{1\over{4}}}\, (1 - 2w)\bar{\Phi}e^{2eV}\Phi
+ {\textstyle{1\over{16}}}\, G^{2},
\end{eqnarray}
with
$\mu'= (0,1,3)$.
Here $G =\bar{D}\bar{\sigma}^{2}DV \sim 2 a^{2}$,
associated with the "reduced" dimension $x^{2}$, is a gauge-invariant
pseudoscalar superfield; 
it acts like a potential for $W_{\alpha}$,
as seen from formulas
 $W_{\alpha} = {1\over{2}}\, (\sigma_{2}\bar{D})_{\alpha}G$
 and $\bar{D}^{2}G^{2} = 8W^{2}$ with 
$(\sigma_{2}\bar{D})_{\alpha}\bar{D}\bar{\sigma}^{2}D  
= -{1\over{2}}\, \bar{D}^{2}D_{\alpha}$.

Note that, on reduction, the coefficient of the $W_{\alpha}$ term 
has changed from $4\kappa \rightarrow  2\kappa$.
(The extra $-2\kappa W_{\alpha}$ term coming from 
$- (\sigma_{2}\bar{D})_{\alpha}{\cal R}^{2}$ was erroneously left out
previously~\cite{KSvor}.) 
Note also that $\int d^{5}z\, {1\over{2}} W^{2} 
= \int d^{7}z\,(- {1\over{4}}\, G^{2})$.
The scale breaking~(\ref{fullbreaking}) thus correctly
incorporates the classical breaking~(\ref{threeDstrcl})  
and, in addition, tells us that the anomaly $-2\delta
\kappa W_{\alpha}$  works to renormalize the FI coupling $\kappa$ in the
quantum supertrace identity,
\begin{equation}
(\sigma_{\mu'}\bar{D})_{\alpha}{\cal R}^{\mu'} 
= 2\kappa_{\rm r} W_{\alpha} +
\bar{D}^{2}D_{\alpha}\Xi_{0}^{\rm (3d)}. 
\label{threeDstr}
\end{equation}
Actually this amount of anomaly was correctly obtained 
by the path-integral method previously~\cite{KSvor}; 
it, however, led to a mismatch of factor 2 for
the renormalization of the FI coupling because of the omission of 
the extra $-2\kappa W_{\alpha}$ term. 
This apparent mismatch is now resolved.

The central charge operator 
$Z= \kappa_{\rm r}\! \oint dx^{k} a^{k}$ is read 
from Eq.~(\ref{threeDstr}).  
Conservation of the current, 
$\partial_{\mu'}{\cal R}^{\mu'} = 0$, directly follows 
from the 4d form of the supertrace,
i.e., Eq.~(\ref{strclassical}) with $4\kappa W_{\alpha} \rightarrow 
(4\kappa -2 \delta \kappa_{\rm r})W_{\alpha}$, 
and this  leads to the algebra~(\ref{scaHiggs}).
The supercharge algebra thus retains the classical form again,
except that the central charge is now renormalized at one loop 
once for all. The vortex spectrum 
$\langle {\rm vor}| P^{0} |{\rm vor}\rangle  = 2\pi n\kappa_{\rm r}/e$
therefore is exact [with $\kappa_{\rm r}$ fixed in terms of $\kappa$ 
as in Eq.~(\ref{kvsv})]. In contrast, the spectrum $m\sim \sqrt{2}ev$
of the elementary excitations is significantly corrected,
with the vacuum expectation value $v$ deviating 
from the classical value so that
\begin{equation}
\kappa_{\rm r}/e =v^{2} - \sqrt{2}ev/(4\pi)+ O(\hbar),
\end{equation}
as discussed earlier~\cite{KSvor}.

\section{Summary and concluding remarks}

In this paper we have studied the effect of quantum fluctuations 
on a sequence of topological excitations that derive from a
classical BPS-saturated soliton or vortex in lower dimensions, and
observed that their spectrum and profile differ critically with the
dimension of spacetime.

In all the examples discussed in the paper the supercharge algebra
retains the classical form although the central charge may
in some cases be modified by short-wavelength quantum fluctuations.
We have shown how to determine such an anomaly by considering
dilatation directly in superspace.

Even though the central-charge operator is explicitly
known, the central charge, given by its expectation value, 
is further governed by long-wavelength quantum fluctuations
and is not necessarily exactly calculable. 
Indeed, for  solitons in a 2d N=1 theory 
the central charge acquires a quantum anomaly at one loop 
but the spectrum is corrected at higher loops as well.
For domain walls in a 3d theory the anomaly simply
contributes to mass renormalization and the central charge is dominated
by the long-wavelength effect that makes the domain-wall spectrum 
only approximately calculable.
For a vortex in a 3d theory the central charge and vortex spectrum 
undergo renormalization at one loop.

The situation changes drastically for 4d theories. 
As we have seen, the spectra of the domain walls and vortices
in 4d theories remain essentially unaffected by quantum fluctuations
(although the profile of the BPS excitations and the spectrum
of elementary excitations are significantly corrected).    
This is a consequence of nonrenormalization of
chiral superpotentials or holomorphy underlying 4d supersymmetry, 
both minimal and extended.
Holomorphy thus has special control
over both short- and long-wavelength quantum fluctuations and 
works to stabilize the BPS-excitation spectrum.   
 This stability is naturally inherited by
topological excitations in theories reduced from 4d theories, 
and in Sec.~IV we have shown how to handle
such a dimensionally-reduced case.

Persistence of BPS saturation in quantum theory is a consequence of the
supercharge algebra rather than dynamics.
Accordingly, once BPS saturation is known, the excitation spectrum is
calculable from the asymptotic excitation profile.   
In view of this, we have emphasized the use of
long-wavelength superfield effective actions
in calculating the central charge and the spectrum in a manifestly
supersymmetric way.

We have used a simple operator  
${\cal D} \equiv m_{\rm r} {\partial\over{\partial m_{\rm r}}}   
+ d_{\lambda} \lambda_{\rm r} {\partial\over{\partial \lambda_{\rm r}}}$ 
[in Eq.~(\ref{operatorD})] to study the effect of dilatation.
The resulting response involves anomalous dimensions such as
$\gamma_{\Phi}$ [in Eq.~(\ref{Cwz})] which refer to the UV divergence.
Our approach is based on renormalized perturbation theory, 
but the outcome is in close resemblance to that of the Wilsonian 
renormalization group (RG) equations~\cite{SV}.
The response equations derived via ${\cal D}$ 
are consistent with those obtained by use of the RG equation and the
Callan-Symanzik (CS) equation and only differ in the choice of bases of
local operators. 
For the RG analysis of the 4d Wess-Zumino model, for example, one may use
\begin{equation}
{\cal D}^{\rm RG} = (1- 2 \gamma^{\rm RG})\, 
m_{\rm r}{\partial\over{\partial m_{\rm r}}} 
- 3 \gamma^{\rm RG} \lambda_{\rm r} 
{\partial\over{\partial \lambda_{\rm r}}} 
+\gamma^{\rm RG}\, N
\end{equation}
(with $N = \int d^{6}z\, \Phi_{\rm r} \delta/\delta \Phi_{\rm r} 
+  \int d^{6}\bar{z}\, \bar{\Phi}_{\rm r} 
\delta/\delta \bar{\Phi}_{\rm r}$ 
being the counting operator~\cite{CDJ})
in Eq.~(\ref{dGr}),
and rewrite Eq.~(\ref{scalebreakingDS}) as 
$\int d^{n}x\, \triangle = - {\cal D}^{\rm RG}S$.
A simple exercise reveals that our $\gamma_{\Phi}$ is related 
to $\gamma^{\rm RG}(\lambda_{\rm r}, m_{\rm r}/\mu)$
so that 
\begin{eqnarray}
\gamma_{\Phi} =  (1- 2z_{m} - 3 z_{\lambda})\, \gamma^{\rm RG}
+  z_{m},
\end{eqnarray}
where $z_{m} = {1\over{2}}\, (m_{\rm r}\partial/\partial m_{\rm r})
\ln Z_{\Phi}(\lambda_{\rm r}, m_{\rm r},\mu, \Lambda)$ and 
$z_{\lambda} = {1\over{2}} 
(\lambda_{\rm r} \partial/\partial \lambda_{\rm r}) \ln Z_{\Phi}$. 
Thus $\gamma_{\Phi}$ deviates from $\gamma^{\rm RG}$ beyond one loop 
and even becomes cutoff ($\Lambda$) dependent. Similarly one finds that
$\gamma_{\Phi} =  (1- 3 z_{\lambda})\, \gamma^{\rm CS}$ in terms of the
anomalous dimension 
$\gamma^{\rm CS}( \lambda_{\rm r})$ of the CS equation.

\acknowledgments

This work was supported in part by a Grant-in-Aid for Scientific Research
from the Ministry of Education of Japan, Science and Culture (Grant No.
17540253).

\appendix

\section{Dilatation in superspace}

In this appendix  we study dilatation directly in superspace.
Let us first consider  N=1 superspace in $n=2,3$
dimensions, discussed in Sec.~III. For a real scalar superfield
$\Phi(z)$ of dimension $d_{\phi}$ the scale transformation is 
written as $\delta_{\rm D} \Phi = (\Omega +d_{\phi})\Phi$
with 
$\Omega = x^{\mu}\partial_{\mu} + {1\over{2}}\bar{\theta}_{\alpha}
\partial/\partial \bar{\theta}_{\alpha}  
= x^{\mu}\partial_{\mu} + {1\over{2}}\bar{\theta}_{\alpha} D_{\alpha}$.
A function $F(\Phi)$ of $\Phi$ then undergoes the change
$\delta_{\rm D} F(\Phi) = \Omega F + d_{\phi} \Phi F'$,
which, on isolating a total divergence 
$\hat{\Omega} = \partial_{\mu} x^{\mu} +
{1\over{2}}\bar{D}_{\alpha} \theta_{\alpha} = \Omega +n-1$, 
leads to the dilatational response 
\begin{equation}
\delta_{\rm D} F(\Phi) \approx 
(1-n) F(\Phi)  + d_{\phi}\Phi F'(\Phi),
\label{dF}
\end{equation}
where $\approx$ means an equality 
under integration $\int dz$ with $dz= d^{n}x d^{2}\theta$;
$F'= \partial F/\partial \Phi$. 
Similarly one can verify that the kinetic term $\int dz\,
(\bar{D}\Phi)D\Phi$ is scale-invariant with $d_{\Phi} = (n-2)/2$.

In two dimensions we set $d_{\Phi}=0$, so that 
$\delta_{\rm D} W(\Phi) \approx - W(\Phi)$.
In three dimensions we set $d_{\phi}=1/2$, obtaining
$\delta_{\rm D} \Phi \approx - {3\over{2}}\, \Phi$ and 
$\delta_{\rm D} W(\Phi) \approx -2 W(\Phi) +{1\over{2}}\, \Phi W'(\Phi)$. 
One can use these relations to isolate a dilatation anomaly
from the effective potentials ${\cal P}_{\rm eff}(\Phi_{\rm c})$ and 
$U_{\rm eff}(\Phi_{\rm c})$, as discussed in a paragraph below
Eqs.~(\ref{Peff}).

In 4d N=1 superspace we use 
$\Omega = x^{\mu}\partial_{\mu} 
+ {1\over{2}}\theta^{\alpha} D_{\alpha}
+ {1\over{2}}\bar{\theta}_{\dot{\alpha}} \bar{D}^{\dot{\alpha}}$
and find for a function of $\Phi_{i} = (\Phi, \bar{\Phi}, V)$ (of
dimension $d_{i}$) the scaling law
\begin{equation}
\delta_{\rm D} F(\Phi_{i}) \approx
(2-n)F  + \sum_{i} d_{i}\Phi_{i}\partial_{\Phi_{i}} F  
\end{equation}
under $\int dz$ with $dz =d^{n}x\, d^{2}\theta d^{2}\bar{\theta}$ 
(and   $n\rightarrow 4$). An interesting example is
$\delta_{\rm D} [\bar{\Phi}\Phi \ln (\bar{\Phi}\Phi)] \approx
2\bar{\Phi}\Phi$, which, for the 4d Wess-Zumino model, 
essentially fixes the one-loop effective action~(\ref{Gammaone}) 
in terms of the anomaly  $-2\gamma_{\Phi} \int d^{8}z\,
\bar{\Phi}\Phi$.
Analogously for a chiral superpotential of the form
$\int\! dz_{\rm c}\, F(\Phi)$ with $dz_{\rm c}\equiv d^{n}xd^{2}\theta$ 
the response $\delta_{\rm D} F(\Phi)$ is given by Eq.~(\ref{dF}).

It is enlightening to work out the scaling law for a vector superfield
$V$ in $n=2,3,4$ dimensions.
Let us suppose that $V$ scales with dimension $d_{V}$,
$\delta_{\rm D}V = (\Omega +d_{V}) V \approx (2-n + d_{V})V$.
Then $W_{\alpha}$ acts with dimension $d_{W}= {3\over{2}} + d_{V}$,
and
$\delta_{\rm D}W^{2} \stackrel{\rm ch}{\approx} (4-n + 2d_{V}) W^{2}$
under $\int dz_{\rm c}$. 
 On the other hand, ${\cal M} = \bar{\Phi}e^{2eV}\Phi$ scales like
$\delta_{\rm D}{\cal M} \approx 2d_{V} eV {\cal M}$ 
with $d_{\Phi} = (n-2)/2$.

There are two natural choices for $d_{V}$.
(i) With  $d_{V} = 0$, $\int dz\, {\cal M}$ is scale invariant.  The
classical variation of the action $S$ in Eq.~(\ref{superHiggs}) is
then written as
\begin{equation}
\delta_{\rm D}^{\rm cl}S = \int dz\, 2(n-2) \kappa V
+ \int dz_{\rm c}\, {1\over{2}}\, (4-n) W^{2}.
\label{Aone}
\end{equation}
(ii)  With  $d_{V} = (n-4)/2$, $\int dz_{\rm c}\, W^{2}$ is scale
invariant, and one finds that
\begin{equation}
\delta_{\rm D}^{\rm cl}S = \int dz\, \Big[ 
2(n-2) \kappa V + (4-n) (\kappa - e{\cal M}) V
\Big].
\label{Atwo}
\end{equation}
These two expressions for $\delta_{\rm D}^{\rm cl}S$ are the same, 
up to classical equations of motion, as explained in Sec.~IV.

A function of $W^{\alpha}W_{\alpha}$, $F(W^{2})$, responds 
to dilatation like Eq.~(\ref{dF}) with $\Phi \rightarrow  W^{2}$ and 
$d_{\Phi} \rightarrow  3 + 2 d_{V}$ under $\int dz_{\rm c}$.
With $d_{V}= (n-4)/2$ one finds, in particular, that 
\begin{equation}
\delta_{\rm D}[W^{2}f(\Phi)]  
\stackrel{\rm ch}{\approx}  d_{\Phi}\Phi f'(\Phi)\, W^{2}
\label{dWWf}
\end{equation}
for a chiral superfield $\Phi$.
This implies that chiral superfields such as $(n-1)^{-1}W^{2}\ln W^{2}$
and
$(2d_{\Phi})^{-1} W^{2}\ln \Phi \Phi_{-}$ are possible components of
a superspace effective action, responsible for the anomaly
$W^{2}$ in Eq.~(\ref{scbHiggs}).


\end{document}